\documentclass[ 
    aps,
    pra,
    twocolumn,
    letterpaper,
    10pt,
    showpacs, 
    showkeys,
    notitlepage,
    amsmath, 
    amssymb,
    floatfix 
]{revtex4-1} 

\usepackage{graphicx}
\usepackage{dcolumn}
\usepackage{bm}
\usepackage{color}
\usepackage{amssymb}
\usepackage{amsmath}
\usepackage{hyperref}

\newcommand{\bel}{\begin{equation}}
\newcommand{\eel}{\end{equation}}
\newcommand{\be}{\begin{equation*}}
\newcommand{\ee}{\end{equation*}}
\newcommand{\bal}{\begin{eqnarray}}
\newcommand{\eal}{\end{eqnarray}}
\newcommand{\ba}{\begin{eqnarray*}}
\newcommand{\ea}{\end{eqnarray*}}

\newcommand{\refeq}[1]{Eq.~(\ref{#1})}
\newcommand{\reffig}[1]{Fig.~\ref{#1}}

\newcommand{\ev}[1]{\langle #1 \rangle}
\newcommand{\ket}[1]{| #1 \rangle}

\newcommand{\+}{^\dagger}

\newcommand{\DD}{\mathcal{D}}

\begin{document} 

\title{Single-photon nonlinearities in two-mode optomechanics}
 
 
\author{P. K\'{o}m\'{a}r$^{1}$}
\author{S. D. Bennett$^{1}$}
\author{K. Stannigel$^{2}$}
\author{S. J. M. Habraken$^{2,4}$}
\author{P. Rabl$^{3}$}
\author{P. Zoller$^{2,4}$}
\author{M. D. Lukin$^{1}$}
 
\affiliation{$^1$Physics Department, Harvard University, Cambridge, 
Massachusetts 02138, USA} \affiliation{$^2$Institute for Quantum Optics and
Quantum Information, Austrian Academy of Sciences, 6020 Innsbruck, Austria}
\affiliation{$^3$Institute of Atomic and Subatomic Physics, TU Wien,
Stadionallee 2, 1020 Wien, Austria}
\affiliation{$^4$Institute for Theoretical Physics,  University of Innsbruck,
6020 Innsbruck, Austria}

\date{\today}

\begin{abstract} 
We present a detailed theoretical analysis of a weakly driven multimode
optomechanical system, in which two optical modes are strongly and
near-resonantly coupled to a single mechanical mode via a three-wave mixing
interaction. We calculate one- and two-time intensity correlations of the two
optical fields and compare them to analogous correlations in atom-cavity
systems.
Nonclassical photon correlations arise when the optomechanical coupling $g$
exceeds the cavity decay rate $\kappa$, and we discuss signatures of one- and
two-photon resonances as well as quantum interference. We also find a long-lived
 correlation that decays slowly with the mechanical decay rate
$\gamma$, reflecting the heralded preparation of a 
single phonon state after
detection of a photon. 
Our results provide insight into the quantum
regime of multimode optomechanics,
with potential applications for quantum
information processing with photons and phonons.
\end{abstract}

\pacs{ 42.50.Wk,  
        42.50.Lc, 
        07.10.Cm  
    }
\maketitle
 
\section{Introduction}
Optomechanical systems (OMSs) involve the interaction between optical and mechanical
modes arising from 
radiation pressure force, canonically in an optical
cavity with a movable mirror \cite{Kippenberg2008,
Marquardt2009, AspelmeyerNJP2008}.
Recent progress in optomechanical (OM) cooling techniques
has been rapid \cite{Metzger2004, Gigan2006, Arcizet2006, Kleckner2006,
Corbitt2007, Schliesser2008,Thompson2008,Wilson2009},
and 
experiments have now demonstrated cooling to the
mechanical ground state
\cite{O'Connell2010,Teufel2011,Chan2011},
OM induced transparency \cite{Weis2010,Safavi-Naeini2011},
and coherent
photon-phonon conversion \cite{Fiore2011,Verhagen2011,Hill2012}.
These developments have
attracted significant interest, and
motivated proposals for applications
exploiting OM interactions at the quantum level,
ranging  from
quantum  
 transducers  
 \cite{Stannigel2010,Safavi-Naeini2011a,Regal2011,Taylor2011}
 and mechanical storage of light \cite{Zhang2003, Akram2010,Chang2011}
 to single-photon sources \cite{Rabl2011} and OM quantum information processing 
 \cite{Stannigel2012,Schmidt2012}.
Significant advantages of OM platforms for these 
applications are
 the possibility for mass production and on-chip integration using
 nanofabrication  technologies, 
 wide tuneability 
 and the versatility of mechanical oscillators 
 to
 couple to a wide variety of quantum systems \cite{Schoelkopf2008}.

The force exerted by a single photon on a macroscopic
object is typically weak; consequently,
experiments have so far 
focused on the regime of strong optical driving, where
the OM interaction
is strongly enhanced but effectively linear
\cite{Groblacher2009,Teufel2011a}
However, recent progress in the design of nanoscale OMSs
\cite{Chan2011,Eichenfield2009, Carmon2007, Ding2011}
and OM experiments in
cold atomic systems 
\cite{Gupta2007, Brennecke2008} 
suggests that the regime of single-photon 
strong coupling, where
the OM coupling strength $g$ exceeds the optical
cavity decay rate 
$\kappa$, is within reach of the
next generation of OM experiments.
In this regime, the inherently nonlinear OM interaction
is significant at the level of single photons and phonons~\cite{Marshall2003,
Ludwig2008, Rabl2011, Nunnenkamp2011}. 
For example, the presence of
a single photon can---via
the mechanical mode---strongly influence or even
block 
the transmission of a second photon, leading to
photon blockade.
This single-photon nonlinearity   was recently
analyzed for canonical OMSs consisting of a single
optical mode coupled to a mechanical mode 
\cite{Rabl2011,Kronwald2012,Liao2012}.
However, with a single optical mode, 
the OM coupling is highly off-resonant, 
leading to a
suppression of effective 
photon-photon interactions by the large mechanical
frequency $\omega_m \gg g$ \cite{Rabl2011}.

In this paper we develop a quantum theory of a weakly driven
\emph{two-mode} OMS
\cite{Miao2009, Stannigel2012,Ludwig2012, BasiriEsfahani2012} 
in which two optical modes are coupled to
a mechanical mode. 
The key advantage of this approach is that
 photons in the two optical modes can be resonantly exchanged by absorbing or
 emitting a phonon via three-mode mixing.
 We extend our earlier results \cite{Stannigel2012}, where we discussed possible
 applications of resonant optomechanics such as single-photon sources and
 quantum gates,
 by exploring one-time and two-time photon correlations 
 of both optical modes.
 Specifically, we find that the photon-photon correlation function of the
 undriven optical mode exhibits delayed bunching for long delay times,
 arising from a heralded single mechanical excitation after detection of a
 photon in the undriven mode.
 Finally, we compare the two-mode OMS
 to the canonical atomic cavity QED system with 
 a similar low-energy level spectrum
 \cite{Carmichael1991, Brecha1999}.
 Despite several similarities we find that, 
 in stark contrast to the atom-cavity system, the OMS studied
 here does {\it not} exhibit nonclassical
 correlations unless the strict strong coupling condition $g >
 \kappa$ is met.
Our results serve as a guideline for OM experiments 
nearing the regime of 
single-photon nonlinearity, and for
potential quantum information processing applications
with photons and phonons.

The remainder of the paper is organized as follows. 
In
Sec.~\ref{sect:Multimode}, we introduce the system and details of the model.
In Sec.~\ref{sect:Averages}, we calculate the equal-time intensity correlation
functions of both  transmitted and  reflected photons, and discuss
signatures of nonclassical photon statistics.
In Sec.~\ref{sect:Delayed_coincidence}, 
we investigate two-time correlation functions
of the transmitted photons, and discuss delayed
coincidence correlations that indicate the heralded
preparation of a single phonon state.
Finally, we provide
a brief outlook on the feasibility of strong OM coupling 
in Sec.~\ref{sect:Experimental values}, 
and 
conclude
in Sec.~\ref{sect:Conclusion} with a summary of our
results.
The Appendix contains details of our analytic model used to
derive several results discussed in the paper.

\section{Multimode optomechanics}
\label{sect:Multimode}

\begin{figure}[htb]
\includegraphics[width=0.45\textwidth]{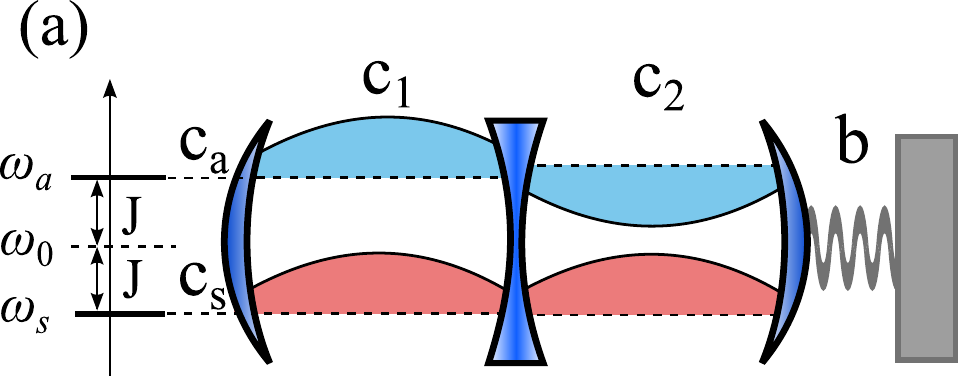} \\[0.5cm]
\includegraphics[width=0.45\textwidth]{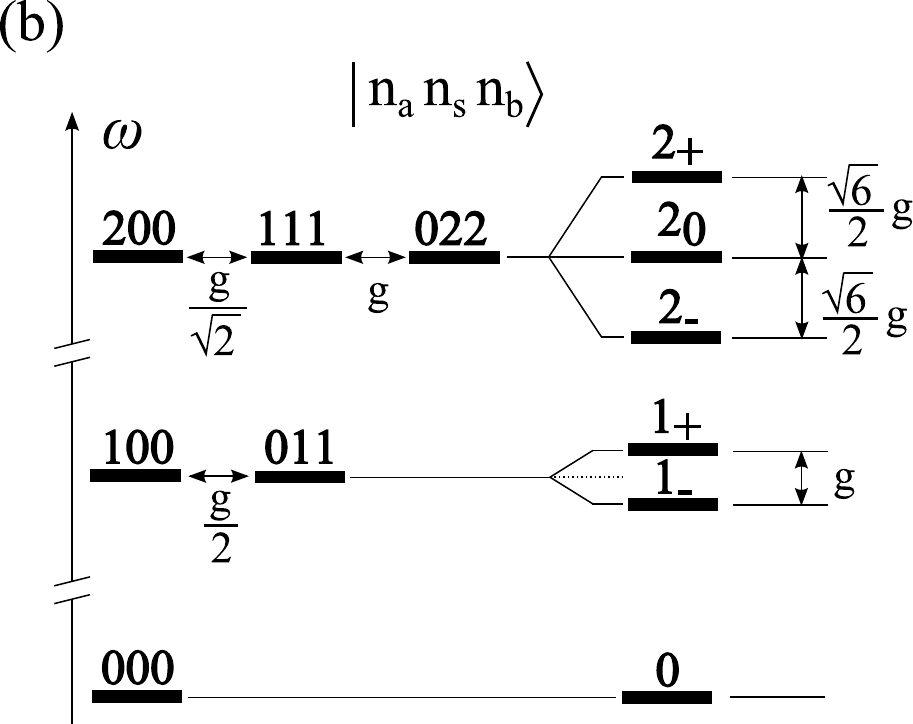}
\caption{ 
  \label{fig:cartoon}
  (a)~Optomechanical system consisting of
  two tunnel-coupled optical cavity modes
  $c_1$ and $c_2$, and a mechanical oscillator $b$ 
  coupled to one of the cavity
  modes by radiation pressure. 
  The coupled optical modes are diagonalized in terms of
  symmetric and antisymmetric modes, $c_s$ and $c_a$.
  (b) Level diagram showing  the  relevant  zero-, one- and two-photon states
  at zero temperature and under the three-mode resonance condition
  $\omega_s - \omega_a = \omega_m$. 
  States are labeled by $\ket{n_a n_s n_b}$ denoting
  the number $n_a$ ($n_s$) of antisymmetric
  (symmetric) photons and the number of phonons $n_b$.
   The optomechanical coupling $g$ splits
   the degeneracy of states  $\ket{n_a n_s n_b}$ and
  $\ket{n_a-1, n_s+1, n_b+1}$.}
\end{figure}

We consider the setup shown schematically in \reffig{fig:cartoon}(a), consisting
of two optical cavities separated by a semitransparent mirror.  The cavity modes
are coupled by photons tunneling through the fixed mirror in the middle, and the mode on the right
couples to the motion of a vibrating endmirror 
through radiation pressure. 
The Hamiltonian describing the system is $(\hbar=1)$ \bel
\begin{split}
\label{eq:Hamiltonian_0}
	H_0 =\; & \omega_0(c_1\+ c_1 + c_2\+ c_2) - J(c_1\+ c_2 + c_1\+ c_2)  \\
	& + \omega_m b\+ b -g(b\+ +	b)c_2\+ c_2  + H_\text{dr}(t),
\end{split}
\eel
where $c_{1,2}$ are annihilation operators for  the two
cavity modes, which we assume to be degenerate with frequency
$\omega_0$, and $J$ is the photon tunneling amplitude through the central
mirror. The motion of the endmirror on the right is described by a single
mechanical mode with annihilation operator $b$ and 
frequency
$\omega_m$, and the parametric
coupling strength $g$ corresponds to the shift of the
cavity frequency due to a single mechanical phonon. Finally,
$H_\text{dr}(t) = \sum_{i=1,2}\left( \Omega_i c_i e^{i\omega_L t}   +
\text{h.c.}\right)$ describes two coherent driving fields of amplitudes
$\Omega_i$ and frequency $\omega_L$, which are applied to the left and right
cavities.

We are interested in a three-mode resonant interaction
in which the two optical modes exchange a photon
by absorbing or emitting a phonon in
the mechanical mode. 
We begin by diagonalizing
the optical part of $H_0$ in the first line of \refeq{eq:Hamiltonian_0}
in terms of the symmetric and antisymmetric combinations
of the optical modes, $c_s = \frac{1}{\sqrt{2}} (c_1 + c_2)$ and
$c_a = \frac{1}{\sqrt{2}}(c_1 - c_2)$, with 
eigenfrequencies  $\omega_{a,s} =
\omega_0 \pm J$. 
In the frame rotating at the laser frequency $\omega_L$ we obtain
\bel
\label{eq:Hamiltonian_tmp}
\begin{split}
	H' = &
	-\Delta c_a\+ c_a  
	- (\Delta + 2J) c_s\+ c_s + \omega_m b^\dag b \\
	&+ \frac{g}{2} 
	\left( c_a\+ c_s +  c_s\+ c_a \right) \left( b + b^\dagger \right)
	+ \sum_{\eta=s,a} \Omega_\eta (c_\eta^\dag +c_\eta),
\end{split}
\eel
where $\Delta = \omega_L - \omega_a$ is the laser detuning from 
the $c_a$ mode,
and $\Omega_{s,a}=(\Omega_1 \pm \Omega_2)/\sqrt{2}$. 
Next, we focus on the case of three-mode resonance,
$\omega_m = 2 J$, and assume that  $\omega_m\gg g,|\Delta|, \Omega_i$.
This  
allows us to make a rotating wave approximation
with respect to the remaining large frequency scale $\omega_m$, and  
in the frame defined by the
transformation
$U = \exp[- i\omega_m t (b\+b - c_s\+c_s)]$,  the Hamiltonian $H'$ simplifies to 
\bel
\label{eq:Hamiltonian_eff}
\begin{split}
	H = -\Delta (c_a\+ c_a+ c_s^\dag c_s) 
	+ \frac{g}{2}(c_a\+ c_s
	b + b\+ c_s\+ c_a)	+ \Omega_a (c_a^\dag + c_a).
\end{split}
\eel
This is the starting point for our analysis discussed below.
Note that the assumptions made in deriving Eq.~\eqref{eq:Hamiltonian_eff} are 
fulfilled in most experimental systems of interest  \cite{Grudinin2010, Zhang2012}, and
$H$  
represents a generic description for
resonant two-mode optomechanics \cite{Miao2009, Dobrindt2010, Cheung2011,
Ludwig2012}.

The nonlinear terms proportional to $g$ 
in  Eq.~\eqref{eq:Hamiltonian_eff} 
describe 
coherent photon exchange between the two optical modes
mediated by
absorption or emission of a phonon. 
The resulting low energy level diagram
is shown in
\reffig{fig:cartoon}(b),
where $\ket{n_a n_s n_b}$ represents a state with
$n_a$ and $n_s$
photons in the $c_a$ and $c_s$ modes, and $n_b$ phonons 
in the mechanical mode.
In the absence of a drive we
diagonalize $H$  in this
few-photon subspace, yielding
the eigenstates 
\bal
	\label{eq:ket0}\ket{0} &=& \ket{000},
	\\
	\label{eq:ket1pm}\ket{1_\pm} &=& \frac{1}{\sqrt{2}}\left(\ket{100} \pm
	\ket{011}\right),
	\\
	\label{eq:ket2pm}\ket{2_\pm} &=& \frac{1}{\sqrt{6}}\left(\ket{200}\pm
	\sqrt{3}\ket{111} + \sqrt{2}\ket{022}\right), 
	\\
	\label{eq:ket20}\ket{2_0} &=& \frac{1}{\sqrt{3}}\left(\sqrt{2}\ket{200} -
	\ket{022}\right).
\eal
Note that in the diagonal basis the weak driving field couples 
all states with photon number differing by one. 
In the following sections we use this low energy Hilbert space
to understand photon correlations in the system.

In addition to the coherent evolution modeled  by the
Hamiltonian $H$, we describe optical and mechanical dissipation
using a master equation for the system density
operator $\rho$,
\bal
\label{eq:master_equation}
	\dot \rho &=& -i[H,\rho] + \kappa \DD[c_a]\rho +
	\kappa \DD[c_s]\rho \\
	&& \qquad + \frac{\gamma}{2}(N_\text{th}+1) \DD[b]\rho +
	\frac{\gamma}{2}N_\text{th}\DD[b\+]\rho,
\eal
where $H$ is  given by \refeq{eq:Hamiltonian_eff}, $2\kappa$ and $\gamma$ are
energy decay rates for the optical and mechanical modes, respectively,
$N_\text{th}$ is the thermal phonon population and  $\DD[\hat o] \rho =
2\hat o \rho \hat o\+ - \hat o\+ \hat o \rho - \rho \hat o\+ o$.
Below we study nonlinear effects at the level
of single photons, both numerically and analytically,
by solving \refeq{eq:master_equation}
approximately in the limit of weak
optical driving, $\Omega \equiv \Omega_a\ll
\kappa$.

\section{Equal-time correlations}
\label{sect:Averages}

\subsection{Average transmission and reflection}

\begin{figure}[tb]
\centering
  \includegraphics[width=0.4\textwidth]{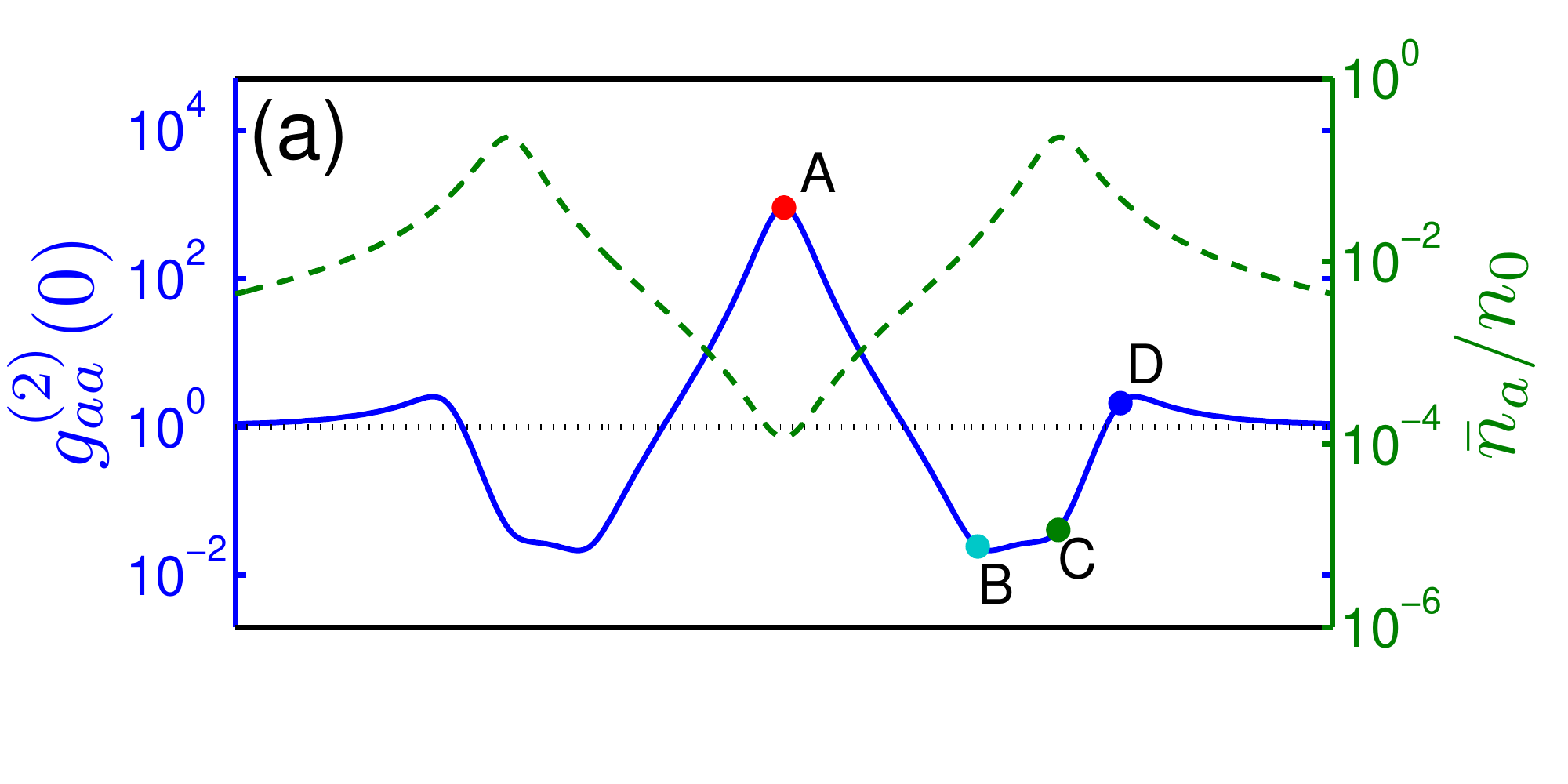}\\[-0.8cm]
  \includegraphics[width=0.4\textwidth]{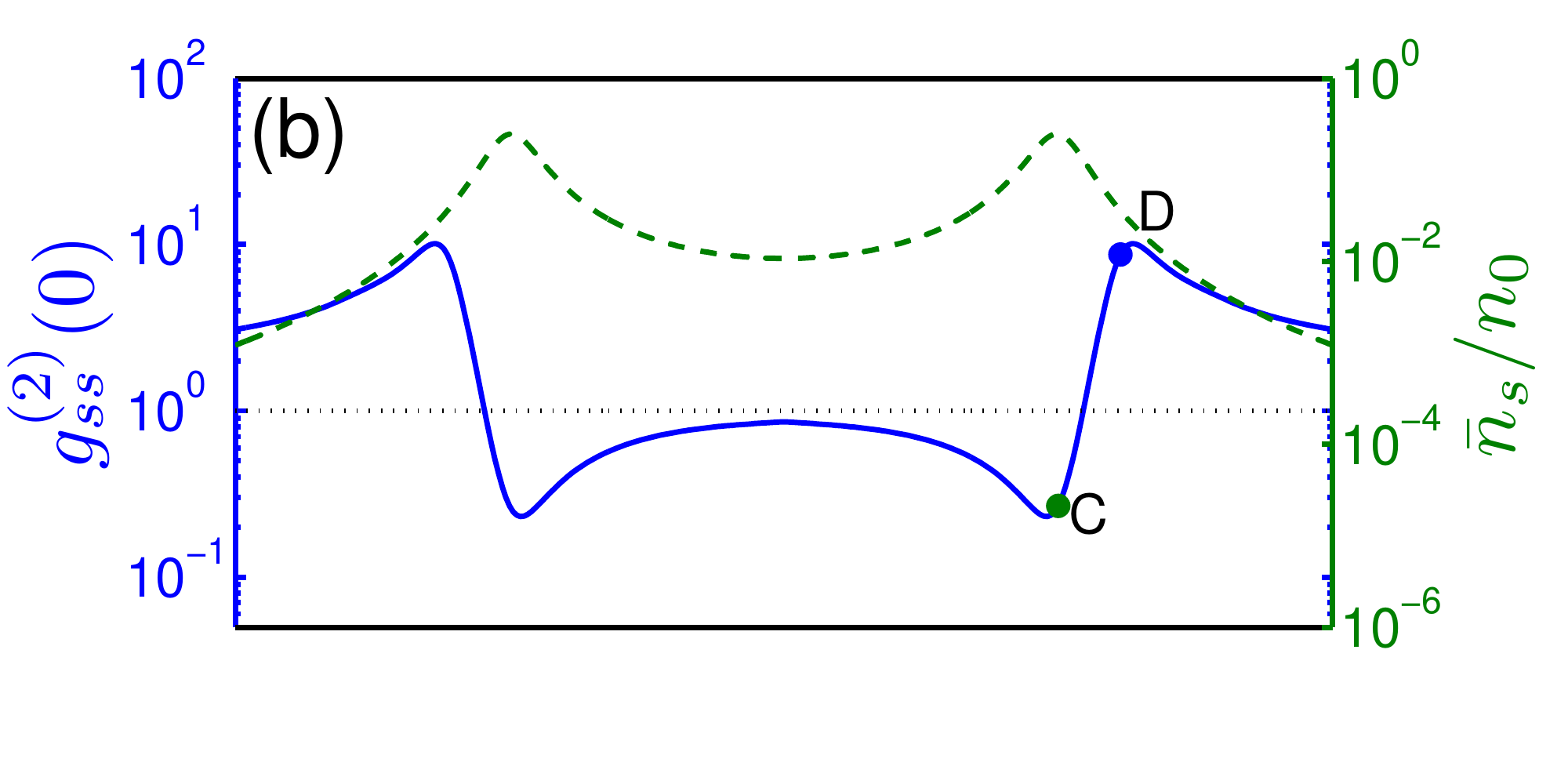}\\[-0.8cm]
  \includegraphics[width=0.4\textwidth]{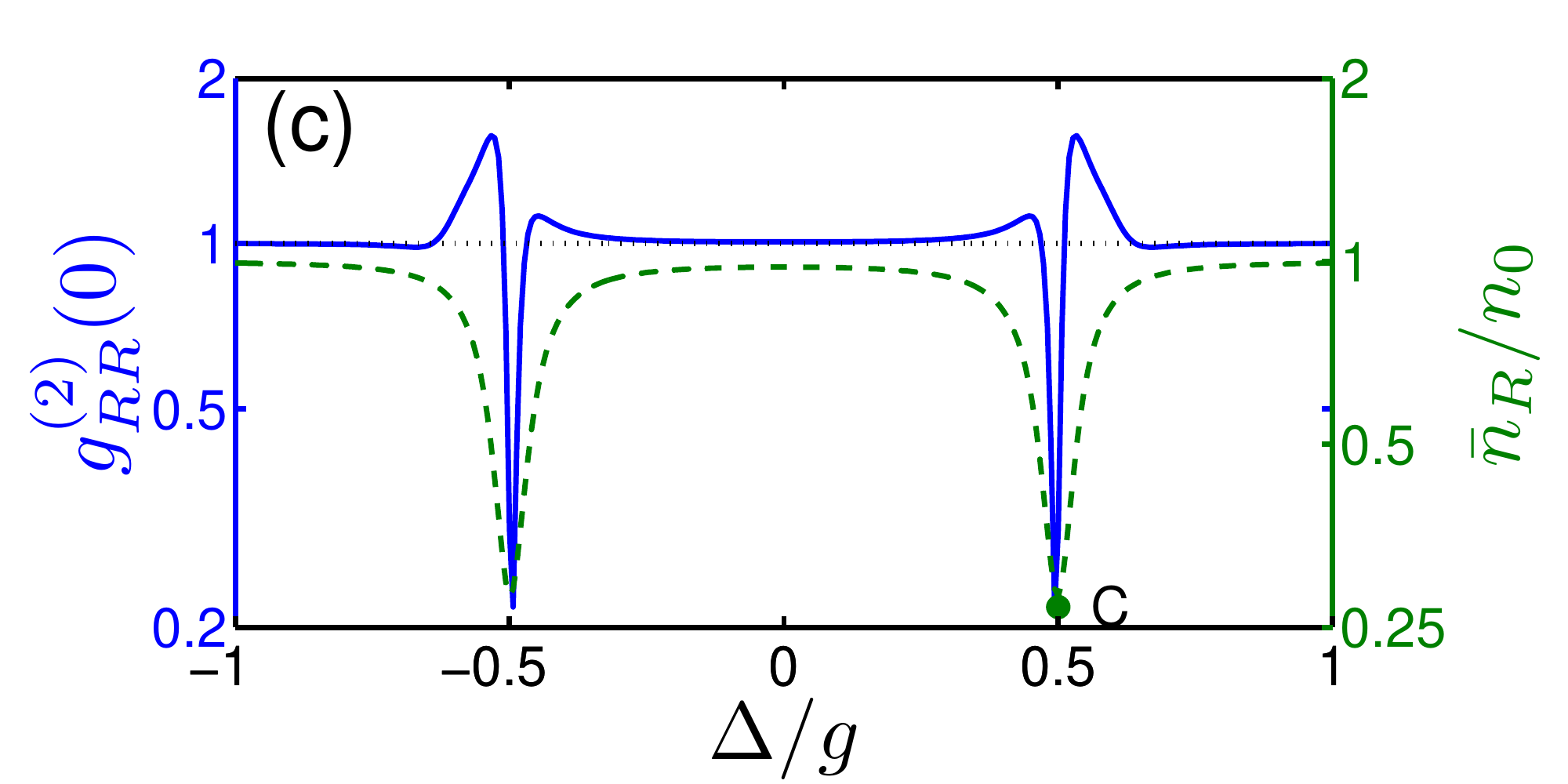}\\[0.3cm]
  \includegraphics[width=0.4\textwidth]{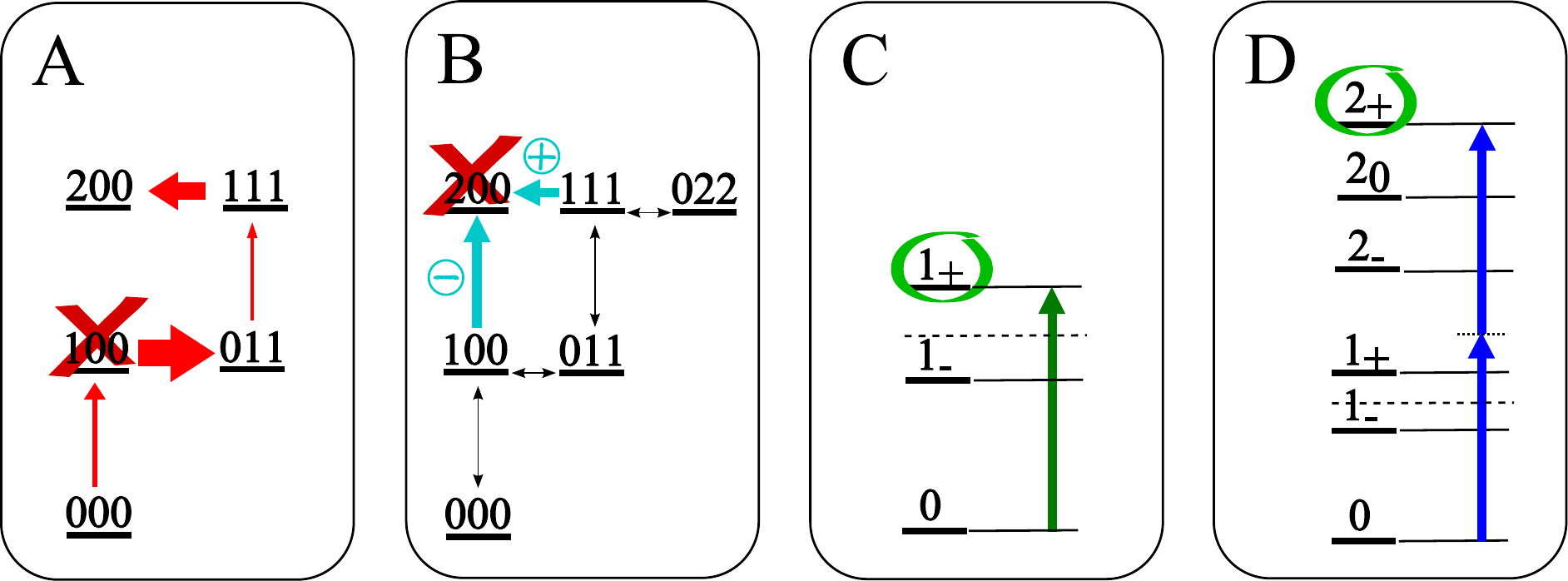}
  \caption{
  \label{fig:spectrum}
  (a) Normalized average photon number (dashed green line) and
  photon-photon correlation function
  (solid blue line) for driven mode $c_a$
  as a function of laser detuning at zero temperature.
  The average photon number is
  normalized by $n_0 = (\Omega/\kappa)^2$.
  (b) Same as in (a) for the undriven mode $c_s$,
  and
  (c) for the reflected field $c_R$.
  In all plots we took
   $g/\kappa = 20$ and $\gamma/\kappa$ = 0.2.
  Dots mark features seen at values of
  detuning $\Delta/g = 0 \text{ (A)}, \frac{1}{\sqrt{8}} \text{ (B)},
  \frac{1}{2} \text{ (C)}$ and  $\frac{\sqrt{6}}{4} \text{ (D)}$.
  Bottom panels A-D illustrate the origin of these features
  as explained in the text. Supressions and enhancements of specific levels is
  indicated by a red X and a green O on top of them, respectively.}
\end{figure}

Before focussing on photon-photon correlations, we first
study
the average transmission through the cavity, which is proportional
to the mean intracavity photon number. In
\reffig{fig:spectrum}(a) and (b) we show the 
intracavity photon number of the
two optical modes (dashed green curves),
\bel
	\label{eq:nAvg}
	\bar n_i = \ev{c_i\+ c_i},
\eel
where $i =a,s$, and angle brackets denote the steady state average.
At $\Delta/g =
\pm \frac{1}{2}$, both transmission curves exhibit a maximum,
indicating that the driving field is in resonance 
with an eigenmode of the system.
The position of these peaks can be  understood from the level diagram shown
in Fig.~\ref{fig:cartoon}(b), which at finite $g$ shows a splitting
of the lowest photonic states into a doublet, $|1_\pm\rangle=(|100\rangle
\pm|011\rangle)/\sqrt{2}$.

In addition to the transmission,
we plot the mean reflected photon number
in  \reffig{fig:spectrum}(c).
As discussed below, the reflected 
photon statistics
can also exhibit signatures of
nonlinearity.
We  calculate properties of the reflected light
using the
annihilation operator 
$c_R = c_a + i\frac{\Omega}{\kappa}$,
obtained from standard input-output relations
for a symmetric two-sided cavity.
The dashed curve in \reffig{fig:spectrum}(c) 
shows the reflected photon number, 
$\bar n_R = \ev{c_R\+ c_R}$. 
At $\Delta/g = \pm \frac{1}{2}$, 
the average reflection has a minimum where the
average transmission has a maximum. 
Note that in contrast to a single cavity, 
even on resonance the transmission probability
is less than unity and the reflection probability
remains finite.

\subsection{Intensity correlations}
\label{sec:onetime_correlation}

To  characterize nonclassical 
photon statistics in the light transmitted though the OMS
we study the equal-time photon-photon correlation
functions,
\bel
	\label{eq:g2(0)}
	g^{(2)}_{ii}(0)
	= \frac{\ev{c_i\+ c_i\+
	c_i  c_i }}{\ev{c_i\+ c_i}^2} ,
\eel
where all operators are evaluated at the same time and $i = a,s, R$. 
A normalized correlation of
$g^{(2)}_{ii}(0)< 1$ indicates photon anti-bunching, and the limit
$g^{(2)}_{ii}(0)\rightarrow 0$ corresponds to the 
complete photon blockade regime
in which two photons never occupy the cavity
at the same time.
The solid curves in \reffig{fig:spectrum} show $g^{(2)}_{aa}(0)$,
$g^{(2)}_{ss}(0)$ and $g^{(2)}_{RR}(0)$ as a function of the laser  detuning and
in the limit of weak driving $\Omega/\kappa \ll1 $.
The most pronounced features of these correlation functions occur at 
$|\Delta|/g = 0, \frac{1}{\sqrt{8}}, \frac{1}{2} \text{ and }\frac{\sqrt{6}}{4}$, 
as marked by
dots A, B, C and D, respectively. 
As we explain in detail in the following analysis, 
we find that the photon bunching at A and
anti-bunching at B are the result of destructive 
quantum interference, while the 
features at points C and D arise from 
one- and two-photon resonances. 


To gain insight into the two photon correlation functions shown in 
\reffig{fig:spectrum} ,
we develop  an approximate analytic model 
for the system 
by considering only the
six levels shown in \reffig{fig:cartoon}(b).
Assuming that the system is initially prepared in $|000\rangle$, these are
the only levels significantly populated by weakly driving the
$c_a$ mode. We make the ansatz \cite{Carmichael1991}
\bel
\label{eq:psi}
\begin{split}
	\ket{\psi} =&\quad A_{000}\ket{000} + A_{100}\ket{100} + A_{011}\ket{011}\\
	&+ A_{200} \ket{200} + A_{111} \ket{111} + A_{022} \ket{022},
\end{split}
\eel
and describe the dynamics by evolving $|\psi\rangle$ under the action of the
non-Hermitian Hamiltonian, $\tilde{H} = H -i\left[\kappa c_a\+ c_a + \kappa
c_s\+ c_s + \frac{\gamma}{2} b\+ b\right]$. 
This approach allows us to evaluate
intensities up to order $\Omega^2$ and two-point correlation up to order
$\Omega^4$, since the neglected quantum jumps
lead to higher order corrections.
By neglecting the typically small mechanical decay rate
$\gamma \ll \kappa$,
the amplitudes in \refeq{eq:psi} then satisfy
\bal
	\label{eq:c000}\dot A_{000} &=& 0,\\[0.2cm]
	\dot A_{100} &=& -i\frac{g}{2} A_{011} -i\Omega A_{000} -
	\tilde\kappa A_{100},\\
	\label{eq:c011}
	\dot A_{011} &=& -i\frac{g}{2} A_{100} - \tilde\kappa A_{011},\\[0.2cm]
	\dot A_{200} &=& -i\frac{g}{\sqrt{2}} A_{111} -i\sqrt{2}\Omega A_{100}
	-2\tilde\kappa A_{200},\\
	\dot A_{111} &=& -i\frac{g}{\sqrt{2}} A_{200} -ig A_{022} - i\Omega A_{011} -
	2\tilde\kappa A_{111},\\
	\label{eq:c022}
	\dot A_{022} &=& -ig A_{111} -2\tilde\kappa A_{022},
\eal
where $\tilde \kappa = \kappa -i\Delta$. 
It is straightforward to solve
Eqs. (\ref{eq:c000}--\ref{eq:c022}) for the steady state amplitudes 
(see Appendix).
To lowest order in $\Omega/\kappa$  the mean
occupation numbers are $\bar n_a=|\bar A_{100}|^2$, $\bar n_s=|\bar A_{011}|^2$
and $\bar n_R=|\bar A_{100}+i\Omega/\kappa|^2$, where $\bar A$ denote
steady state amplitudes. We obtain
\bal
	\label{eq:na}
	\frac{\bar n_a}{n_0} &=&
		\frac{
			\kappa^2\left[R_\kappa(0)\right]^{1/2}
		}
		{
			R_\kappa\left(\frac{g}{2}\right)
		},\\
	\frac{\bar n_s}{n_0} &=&
		\frac{
			g^2\kappa^2
		}
		{
			4R_\kappa\left(\frac{g}{2}\right)
		},\\
	\frac{\bar n_R}{n_0} &\approx &
		\frac{
			\left[R_{\kappa/2}\left(\frac{g}{2}\right)\right]^2
		}
		{
			\left[R_\kappa\left(\frac{g}{2}\right)\right]^2
		},
\eal
where 
$R_K(\omega) = \left[K^2 +
(\Delta-\omega)^2\right]\left[K^2 + (\Delta+\omega)^2\right]$
and
$n_0 = (\Omega/\kappa)^2$.
From the 
factors $R_K(\omega)$ in the denominators  (numerators)
in these expressions, we obtain the positions
of the resonances  (antiresonances) in the 
average intracavity photon numbers, in excellent
agreement with the numerical results
shown in \reffig{fig:spectrum}.
Our six-level model also provides the
equal-time correlations
(see Appendix),
\bal
	g^{(2)}_{aa}(0) &=&
		\frac{
			R_\kappa\left(\frac{g}{\sqrt{8}}\right)
			R_\kappa\left(\frac{g}{2}\right)
		}
		{
			R_\kappa(0)
			R_\kappa\left(\frac{\sqrt{6}}{4}g\right)
		},
		\label{eq:g2aa}
		\\
	g^{(2)}_{ss}(0) &=&
		\frac{ 2\cdot
			R_\kappa\left(\frac{g}{2}\right)
		}
		{
			R_\kappa\left(\frac{\sqrt{6}}{4}g\right)
		},\\
	g^{(2)}_{RR}(0) &\approx &
		\frac{
			R_\kappa\left(\frac{g}{2}\right)
			R_{16\kappa^3/g^2}\left(\frac{g}{2}-\frac{2\kappa^2}{g}\right)
		}
		{
			\left[R_{\kappa/2}\left(\frac{g}{2}\right)\right]^2
		}.
		\label{eq:g2AA}
\eal
Again, these expressions are in agreement with
the features seen in the numerical results in \reffig{fig:spectrum}.
The positions of maxima and minima are
seen directly by the arguments of
the factors $R_K(\omega)$. 
Note that we assumed $g/\kappa \ll 1$ to
obtain the simplified
expressions in Eqs.~(\ref{eq:g2aa}-\ref{eq:g2AA}),
but we retained the 
shift of order $g (\kappa / g)^2$ in
the argument in \refeq{eq:g2AA}
because this shift is larger than the width of 
the antiresonance.



We now discuss each feature
in \reffig{fig:spectrum} in terms of our six-level
model together with the diagonal basis
in Eqs.~(\ref{eq:ket0}-\ref{eq:ket20}).
First, at 
detuning $\Delta = 0$ (point A in \reffig{fig:spectrum})
we see $g^{(2)}_{aa}(0)>1$, indicating bunching.  
This is due to destructive interference that
suppresses
the population in  $\ket{100}$ 
(panel A in \reffig{fig:spectrum}), and
can be understood as the system being
driven into a dark state,
$\ket{d} \propto g \ket{000} - \Omega \ket{011}$, similar to
electromagnetically induced 
transparency (EIT) \cite{Lukin2003, Weis2010}.
In the dark state, $\ket{011}$ remains populated,
allowing transitions to $\ket{111}$
which in turn is strongly coupled to
$\ket{200}$.
The net result is a relative 
suppression of the probability to have one photon
compared to two photons
in the driven mode, 
leading to bunching at $\Delta = 0$.
Second, at detuning $\Delta = g/\sqrt{8}$ 
(point B), 
mode $c_a$ shows anti-bunching
due to a suppressed two-photon probability.
Again, this is due to destructive interference, or
the presence of a dark state in which
$\ket{200}$ remains unpopulated
(panel B).
Third, at detuning $\Delta = \frac{g}{2}$ 
(point C), 
all  modes show anti-bunching.
This is due to
a one-photon  resonant transition
$\ket{0} \rightarrow \ket{1_+}$ 
(panel C).
Finally, at detuning $\Delta = \frac{\sqrt{6}}{4}g$ 
(point D), 
both $c_a$ and $c_s$ show bunching due
to a two-photon resonant transition
$\ket{0} \rightarrow \ket{2_+}$
(panel D).

\subsection{Absence of two-photon resonance at $\Delta = 0$}

At first glance,
the level diagram in \reffig{fig:cartoon} together with
bunching 
in \reffig{fig:spectrum}(a) 
suggest a two-photon resonance
at zero detuning $\Delta = 0$,
where the energy of the
eigenstate $\ket{2_0}$ is equal to the energy of two drive photons.
However, 
as discussed above,
the bunching at $\Delta = 0$ arises 
entirely from
the suppression of a one-photon population;
further, we 
find that the expected two-photon resonance is 
cancelled by  interference.
This can be seen from a second order perturbative 
calculation of the two-photon Rabi frequency $\Omega_{0,2_0}^{(2)}$ for the transition 
$|0\rangle\rightarrow |2_0\rangle$.  
The two-photon state $\ket{2_0}$ can be
populated by the drive 
$H_{\rm dr} = \Omega (c_a^\dag + c_a)$ 
from state $\ket{0}$ 
via two intermediate one-photon
eigenstates, $\ket{1_\pm}$ given by \refeq{eq:ket1pm},
with energies $\omega_{1_\pm} = -\Delta \pm \frac{g}{2}$
in the rotating frame.
The resulting Rabi frequency is
\bel
\label{eq:fgr}
	\Omega_{0,2_0}^{(2)} = \sum_{n=1_-, 1_+}
\frac{\ev{2_0| H_{\rm dr} |n}\ev{n| H_{\rm dr} |0}}{\omega_n},
\eel
which vanishes at $\Delta=0$ 
as a consequence of destructive interference between the two amplitudes. 
The exact cancellation is lifted
by including finite dissipation and the full spectrum;
nonetheless this simple argument shows that
the expected two-photon resonance at $\Delta = 0$
is strongly 
suppressed.

Further evidence of the absence of a
two-photon resonance at $\Delta=0$ is
the lack of bunching in the undriven mode in  
\reffig{fig:spectrum}(b).
If there were a two-photon resonance,
one would expect that bunching
should also occur in the undriven mode,
since the state $\ket{2_0}$ involves both $c_a$
and $c_s$ modes. 
This is indeed the case
at detuning $\Delta = \frac{\sqrt{6}}{4} g$
(see point D in \reffig{fig:spectrum}),
where
both modes show bunching as a result of
two-photon resonance.
In contrast, we see  no bunching in
the undriven mode at $\Delta = 0$. 
This supports our conclusion that the observed
bunching at $\Delta=0$ arises 
from suppression of 
population in $\ket{100}$ due to interference,
as discussed in Section \ref{sec:onetime_correlation},
and not from two-photon resonance.
As discussed above, this interference
does not suppress population in 
$\ket{011}$,
so we do not expect
bunching in the $c_s$ mode from this effect.
\begin{figure}[htb] 
\centering
  \includegraphics[width=0.43\textwidth]{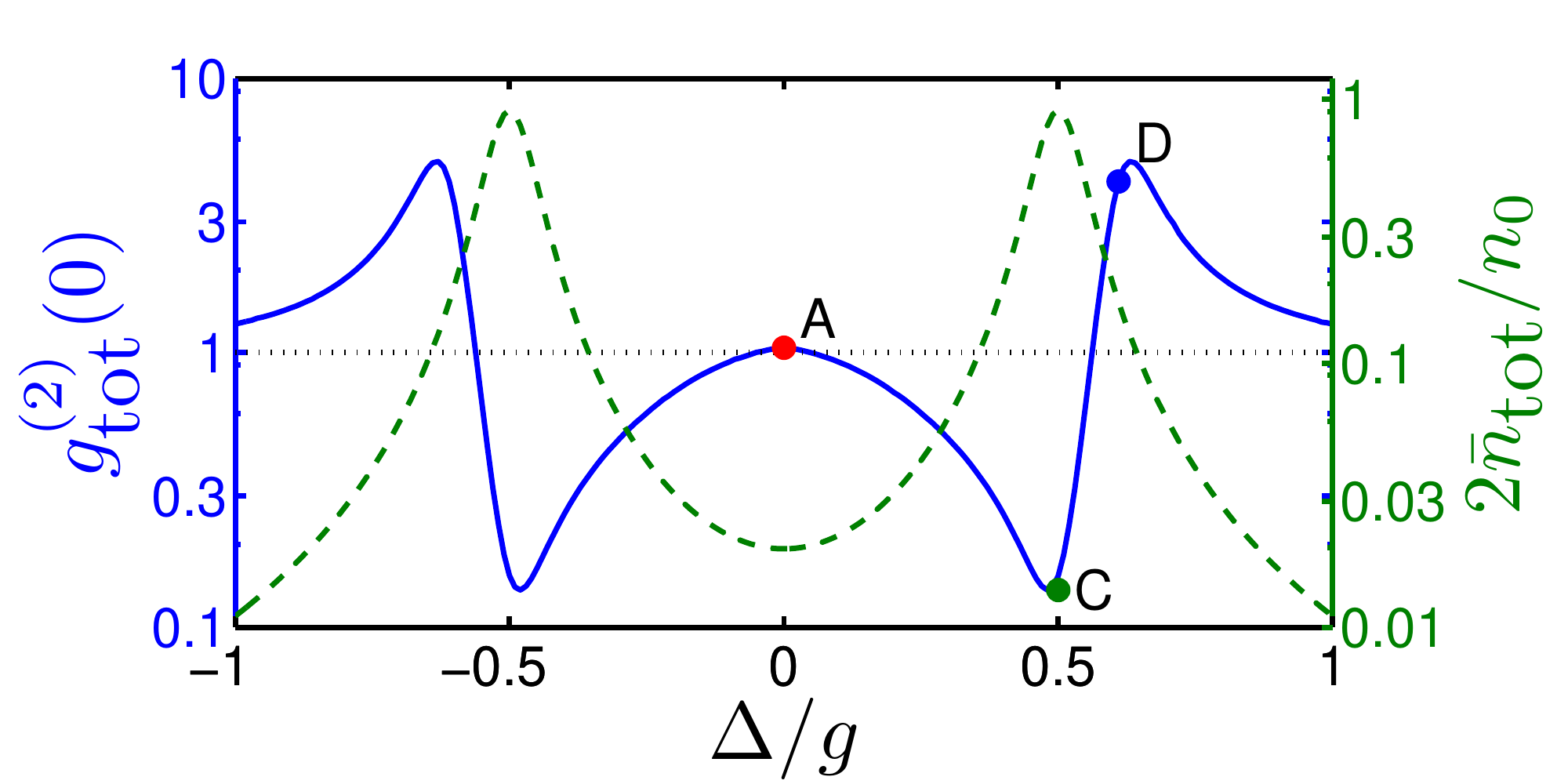}
  \caption{
  \label{fig:populations}
  Average number and intensity correlation
  of total photon number in the coupled OM system.
  Parameters are the same as in \reffig{fig:spectrum},
  and dots mark the same detunings.
  One- and two-photon resonances are
  seen at C and D,
  but we see no bunching in the
  total photon number at 
  $\Delta = 0$ (point A).
  This reflects the lack
  of two-photon resonance due to 
  destructive interference (see \refeq{eq:fgr}).  } 
\end{figure}

Finally, to confirm our intuitive picture we plot the
intensity correlation function, 
$g^{(2)}_\text{tot}(0) =
\ev{n_\text{tot}(n_\text{tot}-1)}/\ev{n_\text{tot}}^2$,
of the total photon number, $n_\text{tot} = n_a +
n_s$,
in the coupled OM system in
\reffig{fig:populations}. 
The probability to find one photon in the combined
cavity is maximal at $\Delta/g=\pm \frac{1}{2}$ 
due to one-photon resonance.
Similarly, we  
observe antibunching at point C
and bunching at point D,
due  to interference and two-photon resonance
respectively,
as discussed 
in Section \ref{sec:onetime_correlation}.
However, we find neither bunching nor antibunching
at $\Delta = 0$, 
demonstrating the absence of a two-photon resonance
despite the fact that $\ket{2_0}$ lies at the twice the
drive frequency.


\subsection{Finite temperature}

So far in our analysis
we have focused on the case where the mechanical
system is prepared in its vibrational 
ground state, $|0_m\rangle$. 
This
condition can be achieved using high frequency resonators 
operated at cryogenic
temperatures \cite{O'Connell2010},
and in the limit of weak driving $\Omega/\kappa\ll1$ such
that optical
heating of the mechanical mode can be neglected. 
In the following we extend our
analytic treatment to the case of finite temperature, 
and show that many of the
nonclassical features are robust even
in the presence of small
but finite thermal occupation of the mechanical mode.

\begin{figure}[tb]
\centering
  \includegraphics[width=0.43\textwidth]{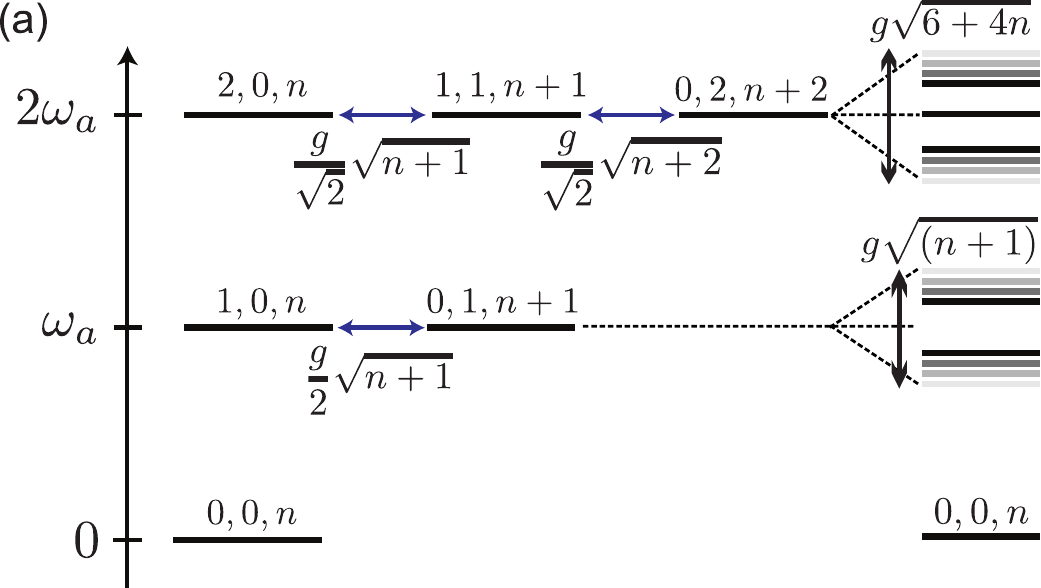}\\
  \includegraphics[width=0.43\textwidth]{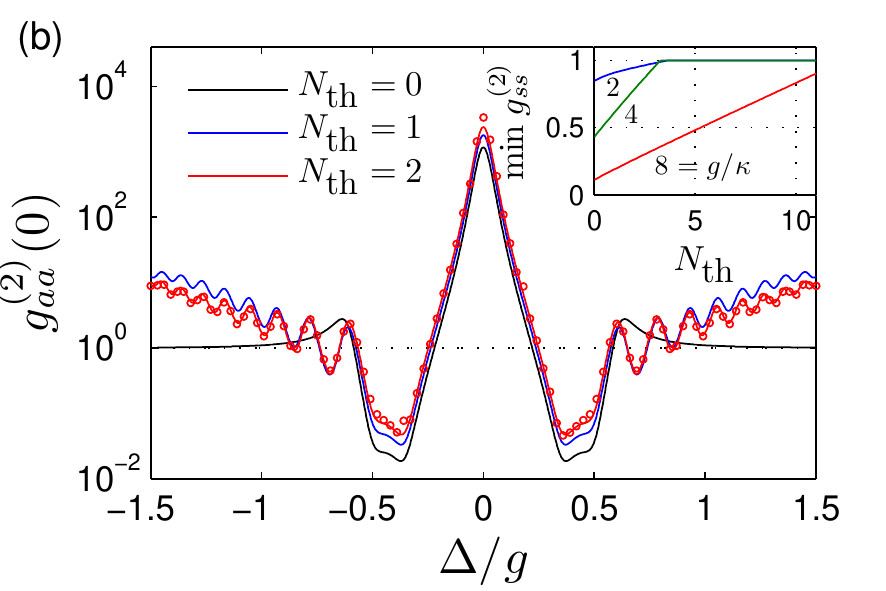}\\
  \includegraphics[width=0.43\textwidth]{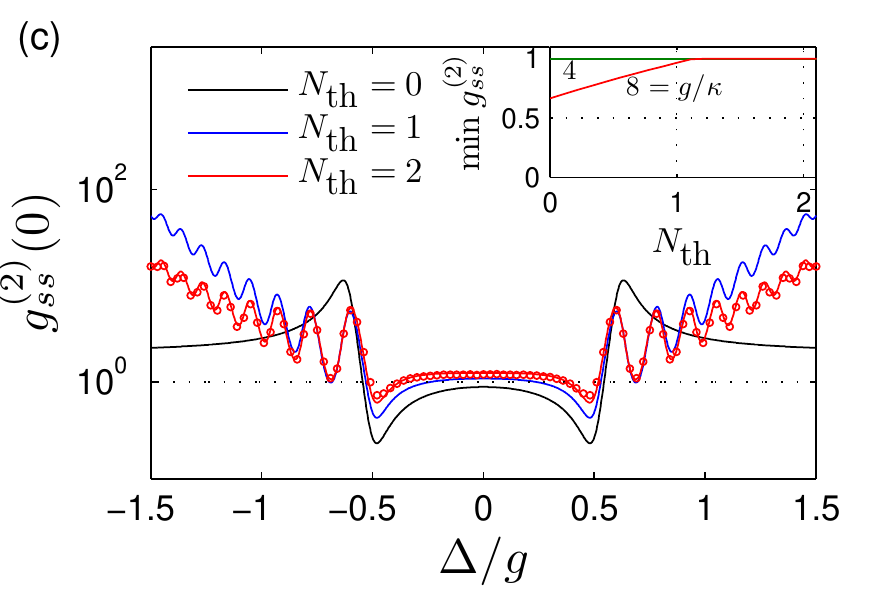}
  \caption{
  \label{fig:thermal_g2}
  Correlation functions for finite temperature. (a) Level diagram showing the
  six states populated by the drive from level $|0,0,n\rangle$ (left), 
  and the associated
  eigenmodes (right) with $n$-dependent splittings.
  (b) Driven mode correlation function $g^{(2)}_{aa}(0)$
  for thermal mechanical occupation $N_{\rm th} = 0,1,2$.
  Solid lines show the analytic calculation with $\gamma \rightarrow 0$,
   and  dots show the full numerical results for $N_{\rm th} = 2$ only.
  The inset shows the minimal $g^{(2)}_{aa}(0)$ as a function of $N_{\rm th}$ 
  for several
  coupling strengths. 
  (c) Same as (b) for the undriven mode correlation function $g^{(2)}_{ss}(0)$. 
  Parameters are $g/\kappa=20$, and (for numerics) $\gamma =  0.001$.
}
\end{figure}

To generalize our previous results we now consider
a finite equilibrium occupation number $N_{\rm th} >0$ of 
the mechanical mode,
but still assume that $\gamma(N_{\rm th}+1) \ll\kappa,g$. 
Within this approximation we proceed
as above, and
make a similar six-level ansatz as in Eq.~\eqref{eq:psi}
for each phonon number $n$,
\bel
\label{eq:psi_thermal}
\begin{split}
	\ket{\psi_n} &=
	 A_{0,0,n}\ket{0,0,n} + A_{1,0,n}\ket{1,0,n} \\
	&+ A_{0,1,n+1}\ket{0,1,n+1} + A_{2,0,n} \ket{2,0,n} \\
	&+ A_{1,1,n+1} \ket{1,1,n+1} 
	+ A_{0,2,n+2}\ket{0,2,n+2},
\end{split}
\eel
 where $\ket{\psi_n}$ includes states up
 to two photons that are connected 
 by the weak drive and coupling $g$,
 starting from the state $|0 0 n\rangle$. 
 As
 shown in Fig.~\ref{fig:thermal_g2} the coupling between the states
 within each six-level subspace
 depends explicitly on the phonon number $n$. 
Following the same approach as above,
the amplitudes in Eq.~\eqref{eq:psi_thermal} evolve according to
\bal
       \dot A_{0,0,n} &=& 0,\\
\dot A_{1,0,n} &=& -i \frac{g}{2} \sqrt{n+1} A_{0,1, n+1} -i\Omega A_{,00,n}
\nonumber\\
&&-\tilde\kappa A_{1,0,n},\\
\dot A_{0,1,n+1} &=& -i\frac{g}{2} \sqrt{n+1} A_{1,0,n} - \tilde\kappa
A_{0,1,n+1},\\
\dot A_{2,0,0} &=& -i g\sqrt{\frac{n+1}{2}} A_{1,1,n+1} -i\sqrt{2}\Omega
A_{1,0,n}
\nonumber\\
&&-2 \tilde\kappa A_{2,0,n},\\
\dot A_{1,1,n+1} &=& -ig\sqrt{\frac{n+1}{2}}  A_{2,0,n} -i
g\sqrt{\frac{n+2}{2}}A_{0,2,n+2} \quad\nonumber\\
&&- i\Omega A_{0,1,n+1} - 2\tilde\kappa A_{1,1,n+1},\\
\dot A_{0,2,n+2} &=& -ig\sqrt{\frac{n+2}{2}} A_{1,1,n+1} -2\tilde\kappa
A_{0,2,n+2}.
\eal
We solve for the steady state amplitudes within each 
subspace  $n$ and average
the result over the initial thermal phonon distribution,
assuming no coupling between subspaces due to
the small phonon relaxation rate.
We obtain the
average photon numbers
\begin{align}
	\bar n_a=\sum_n \zeta_n |\bar A_{1,0,n}|^2,
	\qquad
	\bar n_s=\sum_n \zeta_n |\bar A_{0,1,n+1}|^2,
\end{align}
where $\zeta_n=e^{-\beta \hbar\omega_m n}(1-e^{-\beta \hbar\omega_m })$ and
$\beta^{-1}=k_B T$.
Similarly, the $g^{(2)}_{ii}(0)$ functions are given by 
\begin{align}
	g^{(2)}_{aa}(0)=&2\sum_n\zeta_n |\bar A_{2,0,n}|^2 / \bar n_a^2,\\
	g^{(2)}_{ss}(0)=&2\sum_n\zeta_n |\bar A_{0,2,n+2}|^2 / \bar n_s^2.
\end{align} 
We provide the expressions for the steady state amplitudes
$\bar A_{2,0,n}$ and $\bar A_{0,2,n+2}$ in the
 Appendix.

In \reffig{fig:thermal_g2} we plot the correlation functions,
$g^{(2)}_{aa}(0)$ and $g^{(2)}_{ss}(0)$ for different thermal phonon numbers,
$N_{\rm th}$. 
Solid lines were calculated from the above
analytic approach with $\gamma \rightarrow 0$, 
and we find excellent agreement with
the full numerical results including small but finite
$\gamma$ (dots, shown only for
thermal occupation $N_{\rm th} = 2$).
We see that the zero temperature
features such as antibunching
survive at finite temperature for sufficiently strong coupling~\cite{Stannigel2012}.
In the insets we plot the minimum antibunching
as a function of thermal occupation number for
several ratios $g / \kappa$. 
In addition, for detunings $|\Delta|>g/2$,
a series of new resonances appear in the correlation 
functions, and for small but finite occupation numbers 
we find new antibunching features that
are absent for $N_{\rm th}=0$.
These new features  can be understood from the 
$n$-dependent splitting of the one- and two-photon 
manifolds as indicated in \reffig{fig:thermal_g2}(a). 
For higher temperatures the individual resonances start to overlap, 
and 
we observe an overall increase over a broad region of large positive and negative detunings
due to the cumulative effect of different phonon numbers.

\section{Delayed coincidence and single phonon states}
\label{sect:Delayed_coincidence}

In addition to the equal-time correlations discussed above,
quantum signatures can also be manifested in photon
intensity correlations with a finite time delay.
We now turn to a discussion of  delayed coincidence
characterized by the 
two-time
intensity correlations functions,
\bel
\label{eq:g2_tau}
	g^{(2)}_{ii}(\tau) = \frac{\ev{c_i\+(0) c_i\+ (\tau) c _i(\tau)
	c_i(0)}}{\ev{c_i\+ c_i}^2},
\eel
for both driven and undriven modes, $i = a,s$.
Expressing this correlation in terms of a classical
light intensity $I$,
$g^{(2)}(\tau) = \ev{ I(\tau) I(0) } / \ev{ I }^2$,
and
using the Schwarz inequality, we
obtain the inequalities \cite{Carmichael1991, Brecha1999},
\begin{align}
  \label{eq:classical_criterium}
 	g^{(2)}(\tau) &\leq g^{(2)}(0),	
	\\
	 |g^{(2)}(\tau)-1| & \leq |g^{(2)}(0)-1| .
\end{align}
Similar to the classical inequality
$g^{(2)}(0) > 1$ at zero delay, 
violation of either of these inequalites
at finite delay is
a signature of quantum light. 
We calculate the delayed coincidence correlation
functions
for both the driven and undriven modes.

\subsection{Driven mode}
\begin{figure}[tb] \centering
  \includegraphics[width=0.50\textwidth]{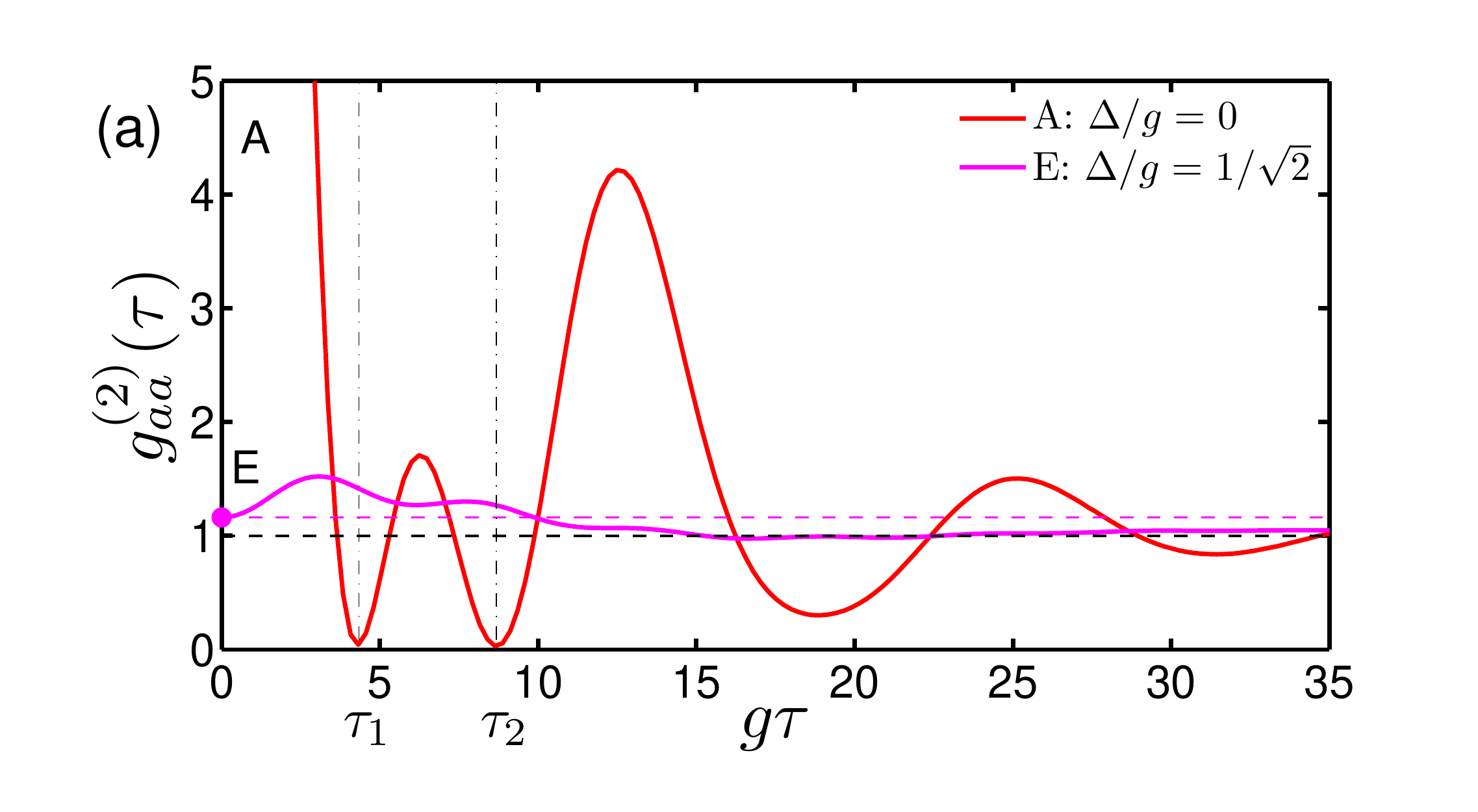}\\[-0.5cm]
  \includegraphics[width=0.50\textwidth]{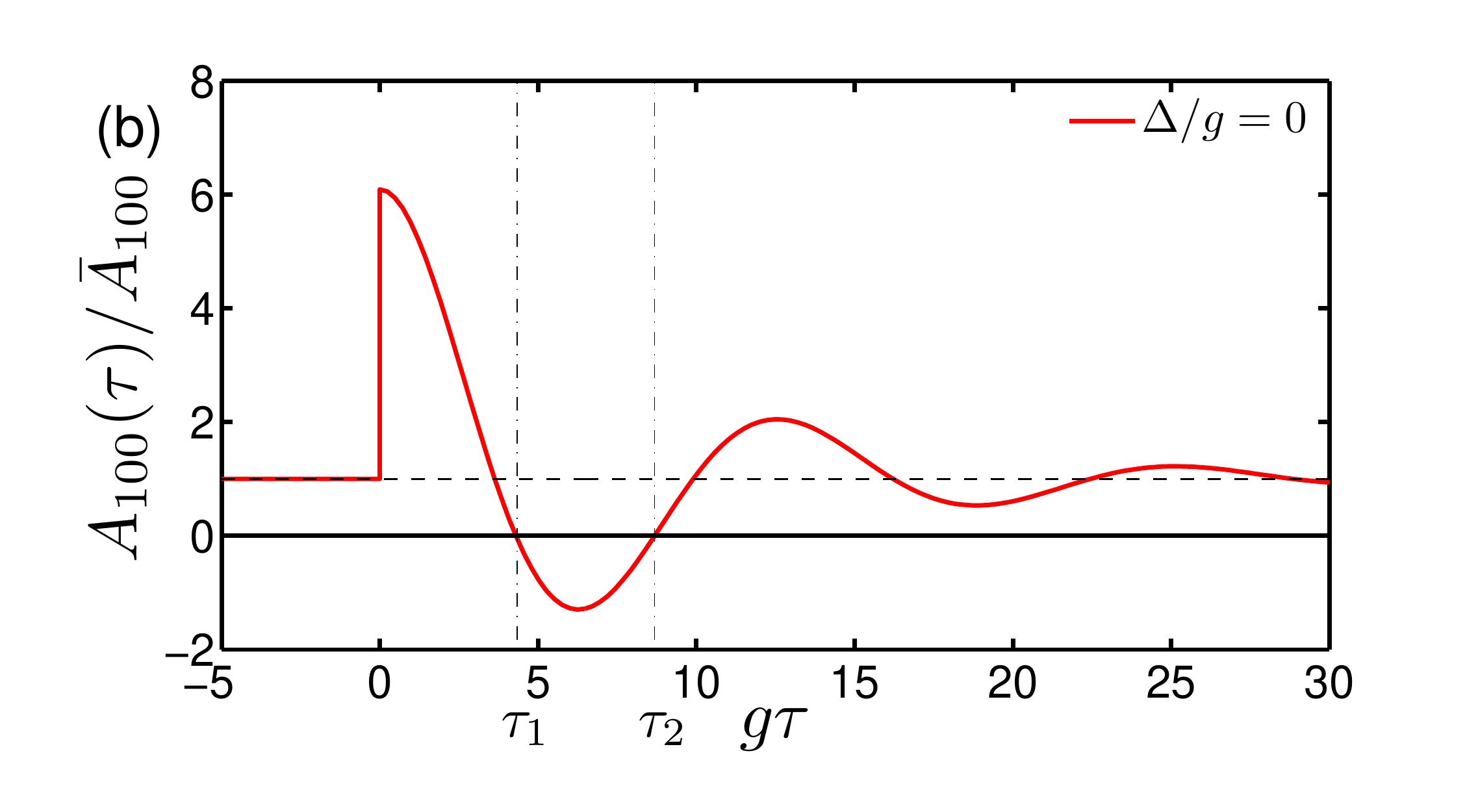}
  \caption{
  \label{fig:g2aatau}
  (a) Finite time delay intensity
  correlation function $g^{(2)}_{aa}(\tau)$  for detunings
  $\Delta/g = 0 \text{ (A)}$ and $\frac{1}{\sqrt{2}}
  \text{ (E)}$.
  Detuning for curve A is the same as marked in \reffig{fig:spectrum},
  while E shows a new effect not seen at equal times.
  Thin dotted line indicates the bound 
  (see \refeq{eq:classical_criterium}) for curve E. 
  (b) Evolution of amplitude $A_{100}$ (normalized by its
  steady state value)
  at detuning A, $\Delta/g = 0$,
  after detecting a driven $c_a$ photon at $\tau = 0$. 
  Vertical dashed lines mark delay times ($\tau_1, \tau_2$) where this amplitude
  vanishes resulting in the vanishing of $g^{(2)}_{aa}(\tau)$
  in (a).
  Parameters are $g/\kappa = 8$ and $\gamma/\kappa$ = 0.02. 
  } 
\end{figure}

The
correlation function $g^{(2)}_{aa}(\tau)$ is shown in \reffig{fig:g2aatau}(a)
for two values of the detuning $\Delta$.
The most striking feature
is the apparent vanishing of  $g^{(2)}_{aa}(\tau)$
at several values of $\tau$ when the
detuning is $\Delta = 0$ (curve A in \reffig{fig:g2aatau}(a)).
These are due to Rabi oscillations at frequency
$g/2$ following the detection of a photon.
This vanishing of the finite delay correlation function
is reminiscent of  wavefunction collapse that
occurs in a cavity containing an 
atomic ensemble \cite{Carmichael1991},
and while its origins are similar, there are important
differences as we now discuss.

We can understand the finite delay intensity
correlations
in terms of the simple six-level model discussed
in the previous section.
We extend
this model to describe finite delay correlations
by considering the effect of photodetection on
the steady state of the system.
Detection of a photon in the driven mode projects
the system onto the conditional state \cite{QuantumNoise},
\bel
	\label{eq:psi^a}
	\ket{\psi^a} = \frac{c_a \ket{\psi}}{||c_a \ket{\psi} ||},
\eel
where $\ket{\psi}$ is given by \refeq{eq:psi} with steady state amplitudes
and $|| \cdot ||$ denotes normalization after the jump.
The conditional state $\ket{\psi^a}$ has an {\it increased}
amplitude $A_{100}$ after the jump
(see jump at $\tau = 0$ in \reffig{fig:g2aatau} (b)).
Following this initial photodetection, the amplitude 
$A_{100}$
subsequently undergoes Rabi oscillations with frequency $g/2$,
and decays back to its steady state at rate $2\kappa$. 
For sufficiently large bunching at zero delay and
strong coupling $g > \kappa$, the  Rabi oscillations of the amplitude
$A_{100}(\tau)$ can cause it to cross zero 
several times before it decays back
to steady state.
As the probability to detect a second photon is dominated by $A_{100}$,
its zeros are responsible for the zeros in the correlation
function $g^{(2)}_{aa}(\tau)$ zero at these delay times.

The zeros in $g^{(2)}(\tau)$ appear similar
to those exhibited  in
a cavity strongly coupled to
an atomic ensemble \cite{Carmichael1991, Rice1988, Brecha1999} or a single atom
\cite{Chang2007}. 
However, in stark contrast to the atomic case,
the zeros in \reffig{fig:g2aatau}(a)
are the result of Rabi oscillations  following
the initial quantum jump.
This is qualitatively different from
the atomic case, where the change in
sign of the relevant amplitude (the analogy of $A_{100}$)
occurs {\it immediately} after the jump itself, and the amplitude
is damped back to steady state at the atomic
decay rate $\Gamma$, without Rabi oscillation. 
As a consequence, the vanishing correlation
function in the atomic case occurs
at a delay set by $\tau_0 \sim \gamma^{-1}\ln C$, 
requiring only strong cooperativity $C = g^2/ \kappa \gamma > 1$
to be visible.
On the other hand,
the zeros in \reffig{fig:g2aatau}(a)
occur at delay times set by  $\tau \sim 1/g$,
requiring strictly strong coupling $g > \kappa$.

Before moving on to correlations of the undriven mode,
we briefly discuss the correlations of the driven mode at the
 other value of  detuning shown in 
\reffig{fig:g2aatau}(a).
 At detuning
$\Delta = \frac{g}{\sqrt{2}}$ (curve E),
which shows bunching at zero time,
$g^{(2)}_{aa}(0) \gtrsim 1$, 
increases  above its initial value at
finite delay.
This is a violation of the classical inequality in
\refeq{eq:classical_criterium}, 
and is an example of ``delayed bunching,"
or an increased probability to detect a second
photon at a finite delay time.
A similar effect was recently studied
in a single mode OMS \cite{Kronwald2012}.
However, like the Rabi oscillations,
the increased correlation function decays back to its steady state
value of 1 on
the timescale of $\kappa^{-1}$.

\subsection{Heralded single phonon states}

\begin{figure}[htb] \centering
  \includegraphics[width=0.50\textwidth]{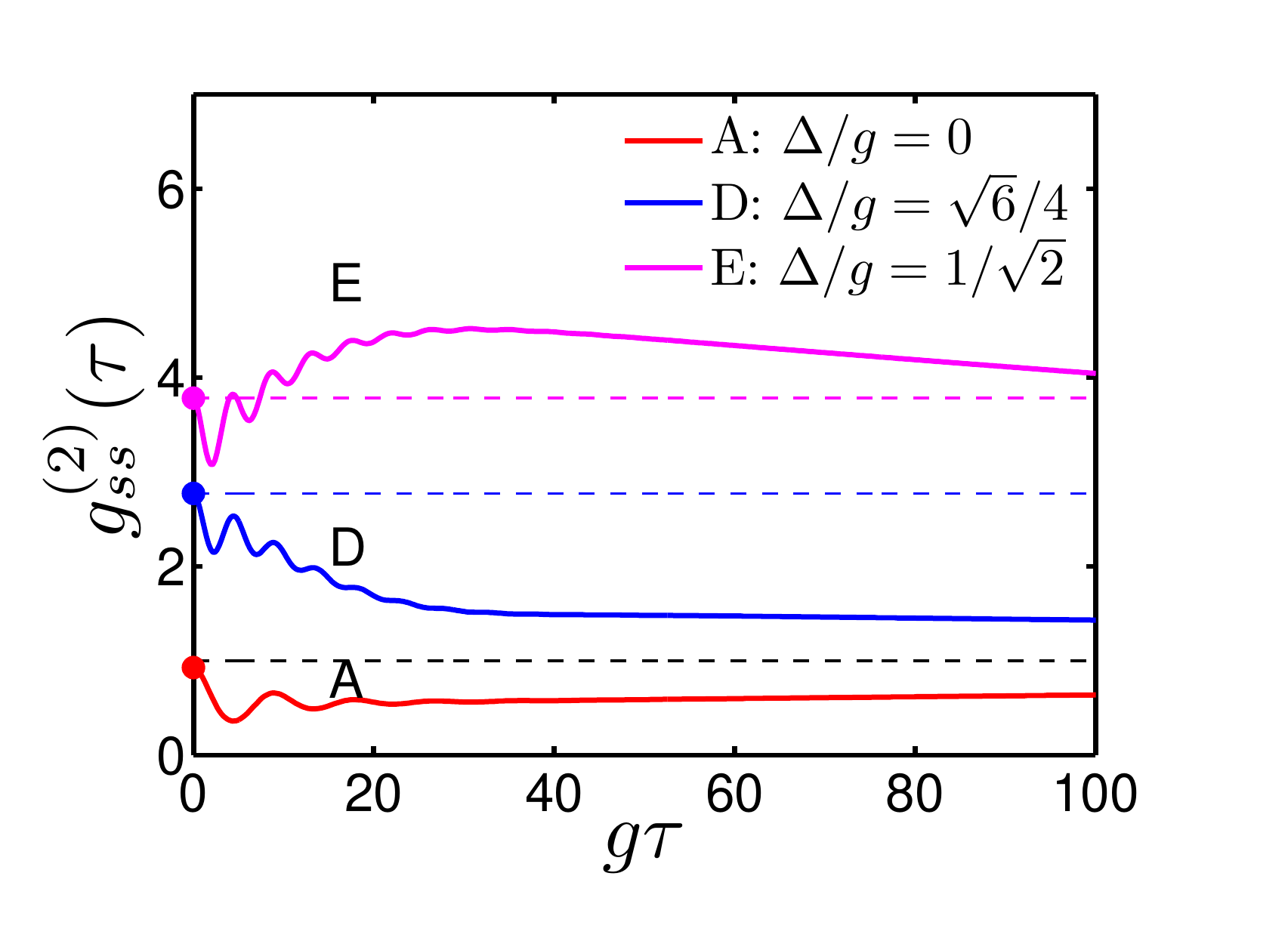}
  \caption{
  \label{fig:g2sstau}
  Finite time delay intensity correlation function $g^{(2)}_{ss}(\tau)$
  for detunings $\Delta/g = 0 \text{ (A)}, \sqrt{6}/4 \text{ (D)}$
  and
  $\frac{1}{\sqrt{2}} \text{ (E)}$.
  Thin dotted lines indicate the classical bounds
  (see  \refeq{eq:classical_criterium}).
  Labels A, D, E correspond to the same detunings
  marked in 
  \reffig{fig:spectrum}
  and
  \reffig{fig:g2aatau}.
  Parameters are $g/\kappa = 8$, and $\gamma/\kappa$ = 0.02.
  }
\end{figure}

We now turn to a discussion of the delayed coincidence
correlations of the undriven mode $c_s$.
We note that correlations of the driven
and undriven modes can be measured separately
provided sufficient frequency resolution, smaller than
  the mechanical frequency.
The correlation function
$g^{(2)}_{ss}(\tau)$
of the undriven mode
is shown in \reffig{fig:g2sstau} for several
values of detuning.
Similar to the driven mode, the
correlation function of the undriven mode
exhibits Rabi oscillations 
that decay on the short optical
timescale
$1/\kappa$.
For detuning $\Delta = 0$ and 
$\Delta/g = \frac{\sqrt{6}}{4}$
(curves A and D in \reffig{fig:g2sstau}),
the correlation $g^{(2)}_{ss}(\tau)$
is described by
our previous six-level 
model of Eqs. (\ref{eq:c000}--\ref{eq:c022}).
However,  at detuning $\Delta = \frac{g}{\sqrt{2}}$
(curve E), we see that
$g^{(2)}_{ss}(\tau)$ has a long
tail that decays
on the much longer mechanical timescale $1/\gamma$.
This is due to the heralded preparation
of a single phonon by detection of a photon in the
undriven mode, as we now discuss.

\begin{figure}[htb]
\centering  
  \includegraphics[width=0.47\textwidth]{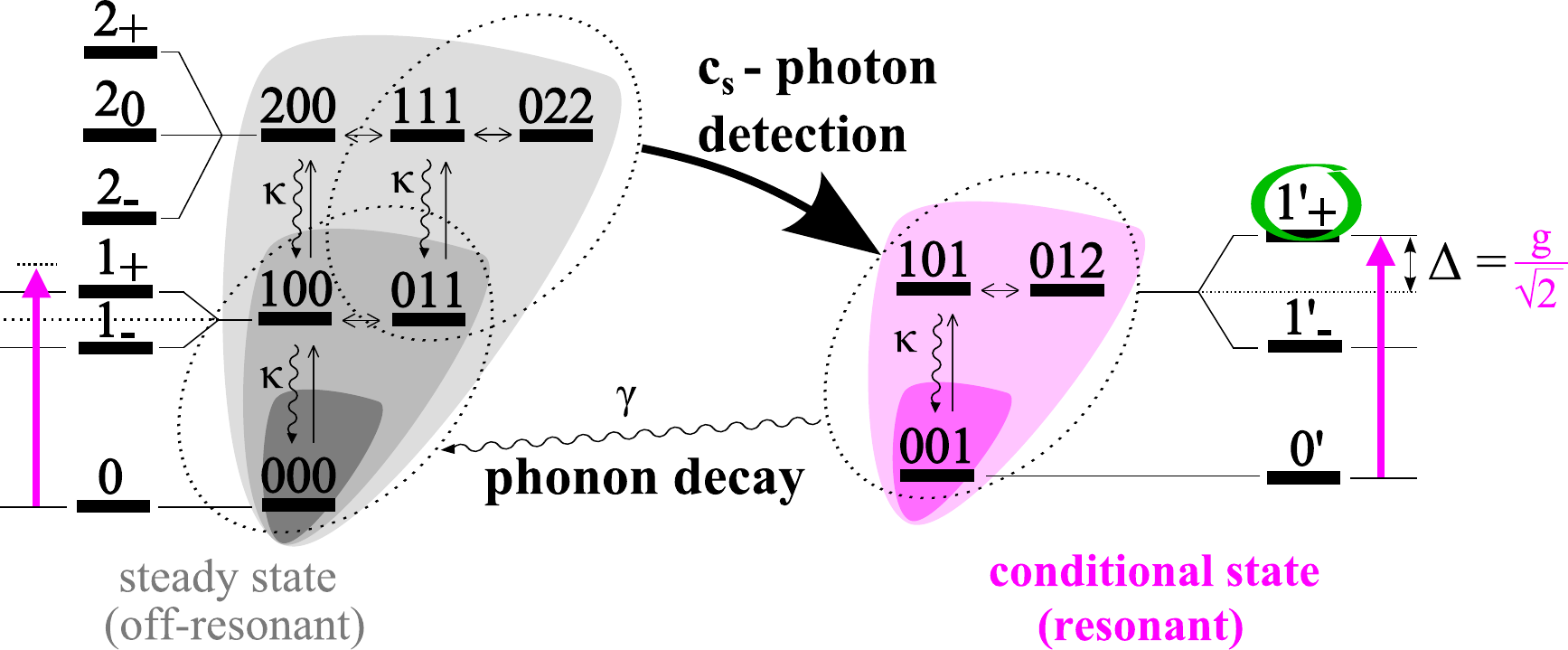}\\
  \caption{
  \label{fig:levels_jumps}
  Effect of detection of a $c_s$ photon
  at detuning E ($\Delta/g = \frac{1}{\sqrt{2}}$).
  In steady state (gray region on left), 
  the drive is far off-resonant.
  However, after detection of a $c_s$ photon the system
  jumps into the conditional subspace (pink region on right).
  Due to the presence of an extra phonon in this subspace,
  the drive is resonant and the probability to detect a second
  photon is much higher than in steady state.
  This increased probability persists
  as long as the extra phonon,
  which decays slowly at rate $\gamma$.
  }
\end{figure}

The increase in delayed coincidence can be understood
by extending the above analytic six-level model to account
for the conditional state of the system after detection
of a photon in the undriven mode.
To do this,
we simply add three additional states to
the six-level ansatz in  \refeq{eq:psi},
\bel
\label{eq:psi9}
	\ket{\psi} = \dots + A_{001}\ket{001} + A_{101}\ket{101} + A_{012}\ket{012},
\eel
since these are the states populated by detection
of a $c_s$ photon from the original six states (see \reffig{fig:levels_jumps}).
Using the same 
approach as before, we obtain the following equations for the
amplitudes,
\bal
	\label{eq:c001}
	\dot A_{001} &\approx& -\frac{\gamma}{2}A_{001},\\[0.2cm]
	\dot A_{101} &=& -i\frac{g}{\sqrt{2}} A_{012} -i\Omega A_{001} -
	\tilde\kappa A_{101},\\
	\label{eq:c012}
	\dot A_{012} &=& -i\frac{g}{\sqrt{2}} A_{101} - \tilde\kappa A_{012},
\eal
where we used $\gamma \ll \kappa$ and
kept the leading term in \refeq{eq:c001}.
We obtain $g^{(2)}_{ss}(\tau)$
by solving these equations for initial
conditions determined by the conditional state $\ket{\psi^s}$
after a quantum jump, 
\bel
	\label{eq:psi^s}
	\ket{\psi^s} = \frac{c_s \ket{\psi}}{||c_s \ket{\psi}||},
\eel
which is a superposition of states $\ket{001}, \ket{101}, \ket{012}$
(see Appendix 
for details), but in the limit of weak driving consist mainly of $|001\rangle$.

Detection of a photon in the undriven mode
implies that the three-wave mixing interaction
converted a photon from the driven mode into
the undriven mode by simultaneously adding a phonon.
The relevant three-level subspace after the jump 
(see \reffig{fig:levels_jumps}) has a similar structure as in the steady state, 
but the presence of an extra phonon
modifies the splitting of the one-photon 
states $|1^\prime_{\pm}\rangle=(|101\rangle \pm |012\rangle)/\sqrt{2}$  to
$\frac{g}{\sqrt{2}}$ 
(instead of $g/2$ without a phonon).
%
This changes
the one-photon resonance condition
for the drive to $\Delta = \frac{g}{\sqrt{2}}$.
Therefore, at this value of the detuning,
the process of
exciting the system and emitting a single $c_s$ photon
is off-resonant;
while after the detection of a first $c_s$ photon
the system is prepared in $|001\rangle$, bringing
it into resonance with the drive.
This {\it enhances}
the probability for subsequent excitation and emission
of a second $c_s$ photon, increasing
the correlation function at finite delay.
The maximum delayed coincidence
occurs after a delay of $\tau \sim 1/\kappa$,
when the photons have reached the
metastable steady state
in the conditional  subspace with one 
extra phonon.
Eventually, the delayed coincidence
returns to its true steady state value
of one
on the timescale
$\tau \sim 1/\gamma$, which is the mechanical decay time of the
state $|001\rangle$. 
Note that the probability to detect a photon from the {\it driven} 
mode
also increases in the conditional state,
so a similar effect is seen in the delayed
cross-correlation function $g^{(2)}_{as}(\tau)$, 
where the photon in the undriven mode is detected first.


\section{Reaching strong coupling}
\label{sect:Experimental values}

The nonclassical correlations predicted in this paper 
require  strong optomechanical coupling, $g > \kappa$, 
as well as sideband resolution, $\omega_m \gg \kappa,\gamma$. 
While the combination of these conditions has not 
yet been demonstrated, 
several experimental efforts 
are currently 
directed at reaching this regime. 
By using micro- and nano-fabricated OMSs such 
as microtoroids or photonic crystal beams, 
high frequency mechanical systems  with 
$\omega_m\approx 50$ MHz - 5 GHz can be combined 
with low loss optical modes, such that the 
condition $\omega_m \gg \kappa \gg \gamma$ is satisfied
\cite{Safavi-Naeini2011a, Schliesser2008, Eichenfield2009}.
At the same time the mechanical system can already 
be prepared close to the quantum ground state by working at cryogenic temperatures. 
In micro-fabricated OMSs, single-photon couplings of 
about $g/\kappa \approx 0.001$ have been demonstrated 
\cite{Eichenfield2009, Ding2011,Verhagen2011}
The largest value to date of $g/\kappa \approx 0.007$ 
has been reached in photonic crystal beam
resonators \cite{Chan2012}, where
colocalized optical and vibrational resonances are
highly confined to maximize coupling
while the surrounding structure is engineered  
to minimize loss.
Conversely, in cold atomic experiments
the effective strong OM coupling regime has been
reached \cite{Purdy2010}, while sideband resolution remains
a challenge \cite{Stamper-kurn}.


There are several existing
proposals for how to meet the challenge
of $g/\kappa > 1$ 
in the photonic crystal beam setup.
First, 
the single-photon optomechanical coupling can 
be increased by making use of
nanoslots in the structure \cite{Robinson2005, Davanc2012} 
to further localize the electric field at 
the position of the mechanical mode.
This could improve $g$ by a factor of 10
\cite{Ludwig2012}.
Second, numerical studies 
suggest that $\kappa$ can be further decreased
by fine tuning the size and position of the slots in
the photonic crystals \cite{Notomi2008,Tanaka2008}. 
Finally, new materials are currently being tested for an 
overall improvement of the OM properties of nano-fabriquated devices \cite{Xiong2012}.
Thus by using these ultrahigh $Q$ photonic crystals or 
similar designs, an increase of $g/\kappa$ by a factor of $\sim 100$ 
is realistic. 
Note that once the strong coupling condition has been achieved, 
the implementation of two or multimode OMSs 
with adjustable tunneling $2J\sim \omega_m$ can be realized via evanescent field
coupling, as has already been demonstrated in the weak coupling regime
\cite{Safavi-Naeini2011a, Eichenfield2009, Grudinin2010}.


\section{Conclusions}
\label{sect:Conclusion}

We have studied  nonclassical intensity correlations in a
driven, near-resonant 
optomechanical system with  one mechanical and two optical modes.
In the regime of strong coupling $g > \kappa$, this system
allows for nonlinear quantum optics through a resonant three-mode 
interaction in which the exchange of two photons is mediated by a phonon.
We have identified several different processes
that can lead to nonclassical antibunching and delayed bunching, 
and we have  
derived a simple analytic model that allows us to 
describe and interpret photon-photon correlations in this 
system both at zero and at finite temperature. 
Our findings will be important as experiments
approach the regime of strong 
OM coupling, and for potential applications of 
OMSs for quantum information processing. 
%
In particular, the long-lived
correlation found for the undriven mode 
raises the intriguing possibility to
exploit such a setup as a quantum memory.
The generation of
heralded single phonons on detection
of a photon from the undriven mode may have implications
for building OM quantum 
repeaters and quantum communication devices.

\acknowledgments 

We are grateful to Pierre Meystre,
Norman Yao 
and Nathalie de Leon 
for enlightening discussions.
This work was supported by
NSF, CUA, DARPA, NSERC, Harvard Purcell Fellowship, 
the Packard Foundation,
the EU network AQUTE, and the Austrian Science
Fund (FWF) through SFB FOQUS and the START grant Y 591-N16.

\appendix*

\section{Analytic model}
\label{sect:App:steady_state}


In this appendix we provide the analytic solutions
used to calculate one- and two-time
correlation functions in steady state.
First, one-time correlations are calculated
from  the steady state solutions of
Eqs.~(\ref{eq:c000}--\ref{eq:c022}).
We set the time derivatives to zero and
solve the equations iteratively,
order by order in the weak drive.
This procedure yields
\bal
	\label{eq:alpha000}\bar A_{000} &\approx& 1,\\
	\label{eq:alpha100}\bar A_{100} &=& -i\alpha \frac{1}{1 + 4x^2},\\
	\label{eq:alpha011}\bar A_{011} &=& -\alpha \frac{2x}{1 + 4x^2},\\
	\label{eq:alpha200}\bar A_{200} &=& -\frac{\alpha^2}{\sqrt{2}}\frac{1 + 2x^2}{(1 +
	4x^2)(1 + 6x^2)},\\
	\label{eq:alpha111}\bar A_{111} &=& i\alpha^2\frac{2x}{(1 + 4x^2)(1 + 6x^2)},\\
	\label{eq:alpha022}\bar A_{022} &=& \alpha^2\frac{4x^2}{(1 + 4x^2)(1 + 6x^2)},
\eal
where $\alpha = \Omega/\tilde\kappa$ ($|\alpha|^2 \ll 1$), $x = g/(4\tilde\kappa)$ and
$\tilde \kappa = \kappa -i\Delta$.  
Using these amplitudes, we can express all
equal time averages.
The mean photon numbers are
\bal
	\bar n_a &=& |\bar A_{100}|^2 ,\\
	\bar n_s &=& |\bar A_{011}|^2 ,\\
	\bar n_R &=& \left|\bar A_{100} + i\frac{\Omega}{\kappa}\right|^2,
\eal 
and the photon-photon correlation functions are
\bal
\label{eq:g2aa0}
	g^{(2)}_{aa}(0) &=& \frac{2|\bar A_{200}|^2}{|\bar A_{100}|^4},\\
\label{eq:g2ss0}
	g^{(2)}_{ss}(0) &= & \frac{2|\bar A_{022}|^2}{|\bar A_{011}|^4},\\
\label{eq:g2AA0}
	g^{(2)}_{RR}(0) &= &
	\frac{\left|-\left(\frac{\Omega}{\kappa}\right)^2+
	2i\frac{\Omega}{\kappa}\bar A_{100} +
	\sqrt{2}\bar A_{200}\right|^2}
	{\left|i\frac{\Omega}{\kappa}+\bar A_{100}\right|^4}.
\eal
To
leading order in $\kappa/g$ these
yield Eqs.~(\ref{eq:na}--\ref{eq:g2AA}).

At finite temperature we calculate steady state amplitudes
within each phonon subspace $n$ similarly
in the ansatz of \refeq{eq:psi_thermal}.
Using the  notation $R_\kappa(\omega)$ introduced 
in Section \ref{sec:onetime_correlation},
the steady state amplitudes
within the subspace with $n$ phonons in the optical
groundstate are
\begin{align}
|\bar A_{10n}|^2=&\frac{\Omega^2\sqrt{R_\kappa(0)}}
{R_{\kappa}\left(\frac{g}{2}\sqrt{n+1}\right)},\\
|\bar A_{01n+1}|^2=&\frac{\Omega^2 g^2 (n+1) }{4
R_{\kappa}\left(\frac{g}{2}\sqrt{n+1}\right)},\\
|\bar A_{20n}|^2 =&\frac{ \Omega^4
R_\kappa(g/\sqrt{8})}{R_\kappa\left(g\sqrt{(2n+1)/8}\right)R_\kappa\left(\frac{g}{2}
\sqrt{n+1}\right)},\\
|\bar A_{02n+2}|^2 =&\frac{ \Omega^4 g^4 (n+1) (n+2)}{32
R_\kappa\left(g\sqrt{(2n+1)/8}\right)R_\kappa\left(\frac{g}{2}
\sqrt{n+1}\right)}.
\end{align}

Two-time correlation functions are calculated similarly,
using the conditional state after a jump
(see Eqs.~(\ref{eq:psi^a}) and (\ref{eq:psi^s}))
as the initial condition.
For example the
unnormalized state after detection of a photon in the
$c_a$ mode is $c_a\ket{\psi} =  \bar A_{100}\ket{000} +
	\sqrt{2}\bar A_{200}\ket{100} + \bar A_{111}\ket{011}$. 
We solve
Eqs.~(\ref{eq:c000}--\ref{eq:c011}) for the amplitudes 
with this state as initial
condition.
The finite delay correlation of the driven mode is
\bel
\begin{split}
	&g^{(2)}_{aa}(\tau) = \frac{|A_{100}(\tau)|^2}{|\bar A_{100}|^4}.
\end{split}
\eel
in good agreement with the numerics.
The correlation of the undriven
mode $g^{(2)}_{ss}(\tau)$ is calculated similarly. 
The unnormalized
state after detection in the $c_s$ mode is  $c_s\ket{\psi} = 
\bar A_{011}\ket{001}  + \bar A_{111}\ket{101} +  \sqrt{2}
\bar A_{022}\ket{012}$.
Using this  as the
initial
condition we solve Eqs.~(\ref{eq:c001}--\ref{eq:c012}) for the amplitudes
in the conditional state. 
In the limit of $\gamma \ll \kappa$ we obtain
\bel
	g^{(2)}_{ss}(\tau) 
	=
	\frac{|A_{012}(\tau)|^2}{|\bar A_{011}|^4}.
\eel

\bibliography{Two_Mode_OM}

\end{document}